# Skybridge-3D-CMOS: A Vertically-Composed Fine-Grained 3D CMOS Integrated Circuit Technology

Mingyu Li[1], Jiajun Shi[1], Mostafizur Rahman[2], Santosh Khasanvis[3], Sachin Bhat[1], Csaba Andras Moritz[1]
[1]Electrical and Computer Engineering, Univ. of Massachusetts, Amherst, USA
[2]Computer Science and Electrical Engineering, Univ. of Missouri, Kansas City, USA
[3]BlueRISC Inc., Amherst, USA

*Abstract*—Parallel and monolithic 3D integration directions offer pathways to realize 3D integrated circuits (ICs) but still lead to layer-by-layer implementations, each functional layer being composed in 2D first. This mindset causes challenging connectivity, routing and layer alignment between layers when connected in 3D, with a routing access that can be even worse than 2D CMOS, which fundamentally limits their potential. To fully exploit the opportunities in the third dimension, we propose Skybridge-3D-CMOS™ (S3DC), a fine-grained 3D integration approach that is directly composed in 3D, utilizing the vertical dimension vs. using a layer-by-layer assembly mindset. S3DC uses a novel wafer fabric creation with direct 3D design and connectivity in the vertical dimension. It builds on a uniform vertical nanowire template that is processed as a single wafer; it incorporates specifically architected structures for realizing devices, circuits, and heat management directly in 3D. Novel 3D interconnect concepts, including within the silicon layers, enable significantly improved routing flexibility in all three dimensions and a high-density 3D design paradigm overall. Intrinsic components for fabric-level 3D heat management are introduced. Extensive bottom-up simulations and experiments have been presented to validate the key fabric-enabling concepts. Evaluation results indicate up to 40x density and 10x performance-per-watt benefits against conventional 16-nm CMOS for the circuits studied; benefits are also at least an order of magnitude beyond what was shown to be possible with other 3D directions.

## I. INTRODUCTION

3D integration is an emerging technology to enable surpassing many of the current limitations in traditional CMOS scaling, including interconnection bottleneck. However, the main research focuses to date, including both parallel integration with Through-Silicon-Vias (TSVs) [1]-[2] and monolithic integration [3]-[4], have been on incremental technology changes based on 2D CMOS. This mindset leads to attempts through die-to-die and layer-to-layer stacking, which fail to realize the real potential in the third dimension in terms of connectivity and density benefits.

In the 3D integrated circuits (ICs) achieved by the parallel and monolithic integration methods, most CMOS Back End of Line conventions are still followed, which leads to very limited connectivity benefits from 3D integration. Parallel integration only adds connectivity using TSVs, which are unlikely to result in significant improvement due to the large TSV pitches and sizes [1]-[2]. As for the state-of-art monolithic 3D integration, transistor-level monolithic 3D IC has the most fine-grained vertical connections to date thanks to the small monolithic inter-tier vias (MIVs) in cells, while still uses conventional CMOS interconnections for inter-cell connections despite the shrinking cell footprints, which leads to various issues including pin and routing congestions [5]. Specifically, in transistor-level monolithic 3D, routing congestion is caused by reduced pin access on the input/output metal port of each standard cell. While a typical 14nm FinFET based 2D cell has at least 6 pin access points, a 3D cell may have only 3-4 due to its reduced footprint and the area occupied by MIVs. Gate-level monolithic 3D IC has even less fine-grained vertical connections than transistor-level 3D IC [5].

In addition to the connectivity challenge, even in the most fine-grained 3D solution, namely, monolithic 3D integration, the maximum number of vertically-stacked transistors is limited to the number of active silicon layers. This limitation prevents us from achieving higher circuit density and in turn leads to limited connectivity benefits, in addition to stringent registration and alignment requirements. Moreover, heat management is also challenging because of the higher power density and longer heat dissipating paths in such 3D ICs. All these limitations encouraged us to pursue a 3D integration technology with a truly fine granularity in all dimensions, including vertical, which addresses these issues in conjunction.

In this paper we present Skybridge-3D-CMOS (S3DC), a fine-grained 3D IC fabric technology that builds on uniform vertical nanowire templates and utilizes novel assembly, interconnect and heat extraction structures designed with a 3D mindset. Overall fabric direction enables higher density and connectivity vs. monolithic 3D integration by (i) a vertically oriented design-framework enabling unique connectivity approaches avoiding routing issues; (ii) integrated wafer processing vs. a layer-by-layer approach; and (iii) intrinsic heat extraction architecture. In Section II we introduce how the afore-mentioned key challenges are solved with the proposed S3DC fabric concept, components and circuits. In Section III the fabric is evaluated for several circuits we designed in 3D to quantify its benefits vs. state-of-the-art CMOS. In Section IV the S3DC manufacturing pathway and experimental progresses are discussed. In Section V we compare the S3DC fabric with other 3D technologies and highlight how the 3D integration challenges are addressed.

## II. S3DC FINE-GRAINED-3D-ENABLING FEARURES AND CIRCUIT DESIGNS

S3DC fabric is designed to realize a new kind of fine-grained 3D integration by co-envisioning its fabrication

This work has been supported by National Science Foundation (NSF) grant 1407906, and Center for Hierarchical Manufacturing (CHM, NSF DMI-0531171) at UMass Amherst.

process, assembly and fabric components. It is based on a *uniform* vertical pre-doped nanowire template, which is shown in Figure I. Then the template is functionalized by multi-layer selective material deposition. All the S3DC component structures are designed to fit this pathway for 3D manufacturability. The functionality of these components and structures depends on their material types and geometries. This manufacturing pathway is similar with another non CMOS based 3D IC fabric we proposed earlier, called Skybridge [6]-[9] and with a device that we experimentally demonstrated at a sub 30-nm scale [10]. The detailed manufacturing pathway is covered in Section IV.

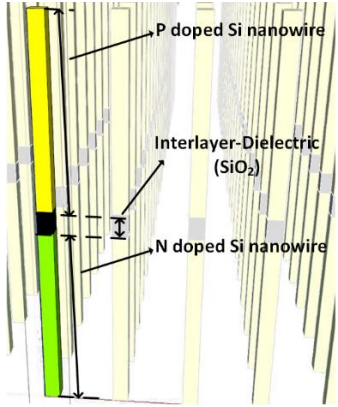

Figure I. Uniform vertical nanowire template

### A. S3DC Fabric Components

S3DC technology greatly improves the connectivity thanks to its innovative 3D interconnect structures as well as higher intrinsic circuit density. In contrast to traditional CMOS and parallel / sequential integration where interconnections are mostly made through horizontal metal wires in many metal layers above the transistor layers, S3DC fabric features a true 3D interconnection framework with specifically architected 3D interconnection structures as shown in Figure II: i) Bridges provide horizontal connections between adjacent nanowires; ii) Coaxial Routing structures carry signals vertically along nanowires; iii) nanowires can be used for vertical routing since they are heavily-doped and silicided and thus have good conductivity; iv) Skybridge-Interlayer-Connection (SB-ILC) provides good Ohmic contact between p- and n-doped regions of a nanowire, which is designed with materials chosen based on the required work function (detailed structure and chosen materials shown and explained in Figure II(B)). This framework provides very high connectivity due to its routing flexibility in all the three dimensions.

Uniform Vertical Gate-All-Around (V-GAA) Junctionless transistor acts as the active device in S3DC technology; an n-type transistor structure is shown in Figure III. The source, channel, drain regions of these transistors are based on the heavily-doped vertical nanowires, and gate electrodes and dielectric layers are selectively deposited surrounding the nanowire. The device behavior is modulated by the work function difference between gate electrodes and the channels. The concept of this type of transistors has been well researched [11], and also experimentally demonstrated in our group [10]. Although Junctionless device is often considered as providing lower on-current and somewhat worse device-level performance, in S3DC it is a part of a true fine-grained 3D integration solution, with dense connections and design, that overall yields higher performance at the fabric-level despite a somewhat sub-optimal device vs state-of-the-art FinFETs.

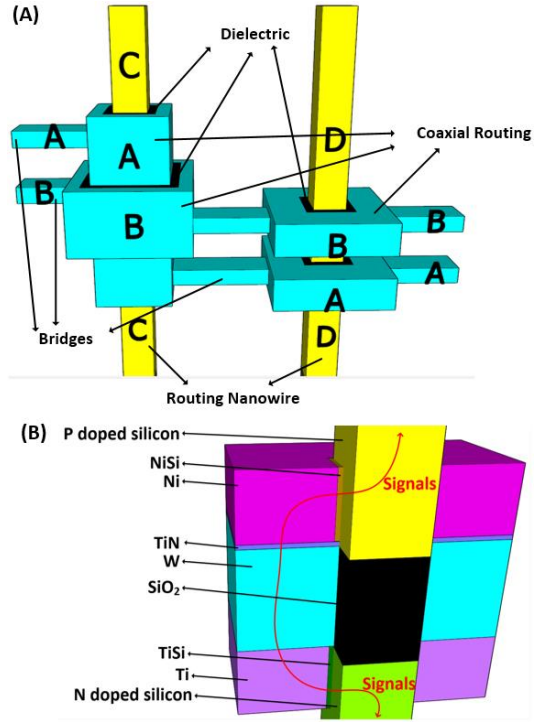

Figure II. S3DC interconnection components: A). 3D connections within one doping layer realized by Bridges, Coaxial Routings, and routing nanowires; four signals A, B, C, D are carried in this example; B). SB-ILC allows routing between various doping layers without MIVs; Ni and Ti are chosen for providing good Ohmic contact with p- and n-doped silicided Si, respectively; TiN acts as diffusion barrier between Ni and W.

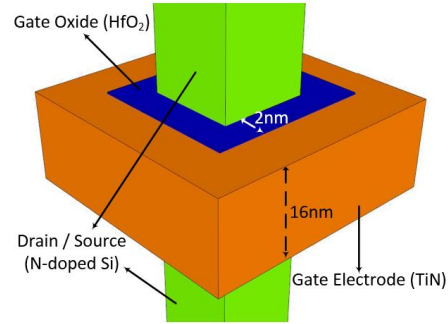

Figure III. An n-type V-GAA Junctionless transistor in 16-nm S3DC technology

### B. S3DC Circuits

S3DC circuits are implemented following the static CMOS circuit style. A three-input NAND gate is shown in Figure IV as an example of a logic-implementing standard cell. Three parallel p-type transistors on the top p-doped region act as the pull-up network and three serial n-type transistors at the bottom as the pull-down network. SB-ILC connects the pull-up and pull-down circuits to generate the output signal, which is conducted out by Bridges. The functionality of the circuits have been verified through detailed physical simulation; the details are covered in the S3DC evaluation section.

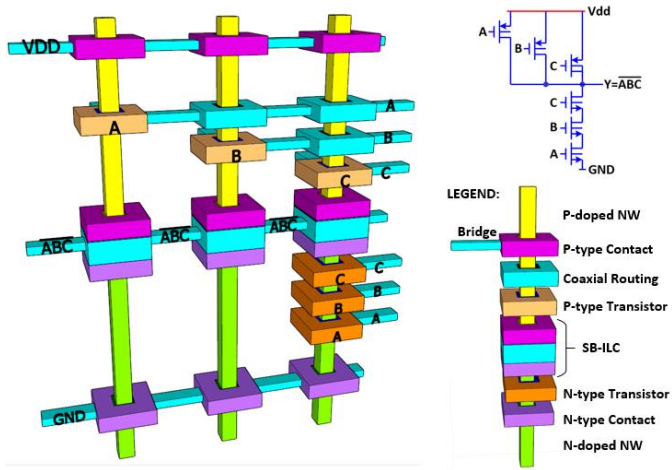

Figure IV. S3DC 3-in NAND gate layout (dielectric for isolation between components and structural support not shown).

Figure V shows the S3DC 6-T SRAM design, which conforms to the S3DC integration requirements. Unlike conventional 6-T CMOS SRAM requiring device sizing, S3DC SRAM uses uniform transistors for better 3D manufacturability.

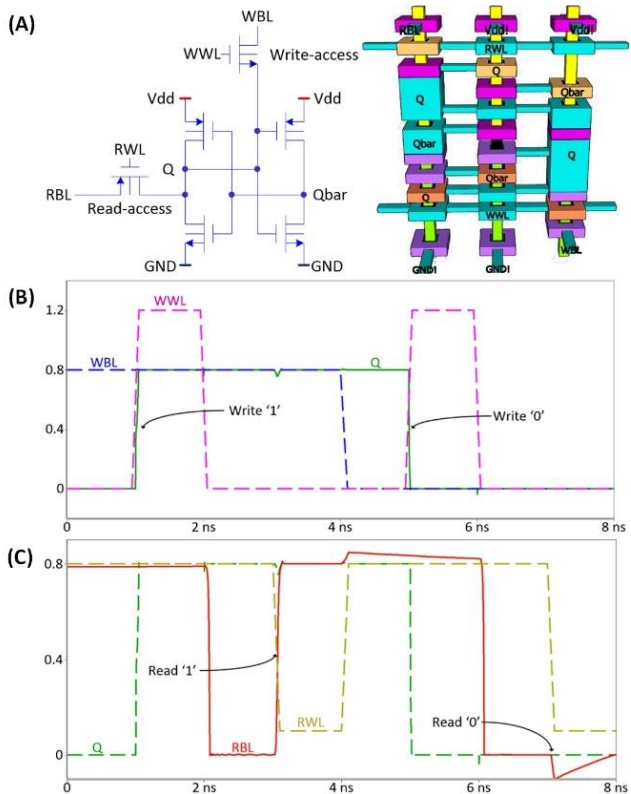

Figure V. S3DC SRAM: A). 6-T S3DC SRAM cell design; B). write operations; C). read operations.

S3DC SRAM cell stores value with cross-coupled inverters, and provides separate read and write accessibility controls with two pass transistors. During the read operation, Read-Bit-Line (RBL) is initialized to "0" and left floating, then Read-Word-Line (RWL) driver generates 0.1V to weakly turn on the p-type read access transistor, allowing RBL to read the stored value without flipping it; during the write operation, Write-Bit-Line (WBL) is driven to "0" or "1"by write driver, then Write-Word-Line (WWL) driver generates 1.2V to strongly turn on the n-type write access transistor, making WBL overpower the feedback inverter and write into the SRAM cell. Using new voltage levels on word-lines only requires small overhead from customizing row circuitry, the word-line drivers. Read stability and writability, however, are ensured without requiring precise device sizing, or increased control and routing complexity. This mindset of enhancing noise margin with multiple word-line voltage levels has been proved to be effective in applications such as Wordline Underdrive [12].

*C. S3DC Heat Management*

In order to tackle the critical 3D challenge which is heat management, S3DC technology employs intrinsic fabric-level heat extraction structures supporting circuits instead of following the conventional mindset which only incorporates heat management considerations during the IC design cycle. In S3DC, we envision that heat extraction is added during the circuit synthesis / design stage through CAD. Heat extraction components, including Heat Extraction Junctions (HEJs) and Heat Dissipating Power Pillars (HDPPs), are specifically designed for S3DC following design principles as in [7] to provide distributed heat dissipation paths. HEJs are specialized junctions that extract heat from the hot spot on a nanowire without affecting circuit function; the extracted heat is carried by Heat Extraction Bridges and is dissipated to the bulk silicon substrate through HDPPs, which are larger dimension pillars.

## III. S3DC FABRIC EVALUATION

In this section we show the S3DC fabric evaluation following extensive simulation methodologies. Benefits from fabric-level heat management features were evaluated first. Then power, performance and density benefits have been quantified and compared.

*A. Thermal Management Evaluation*

The S3DC's thermal management was evaluated with analogous analysis in the electrical domain [13]. Equivalent thermal resistance models for transistors and logic-implementing nanowires following similar principles in [7] have been developed. Next, we built benchmark circuits in scenarios when two gates with various numbers of transistors are stacked on one nanowire, and completed HSPICE simulations for worst-case heat dissipating scenarios when the transistors generate most total heat. Hot spot temperatures were recorded as the metric for evaluating heat management.

TABLE I. WORST-CASE HOT SPOT TEMPERATURE

|  | Inverter | 2-in NAND | 3-in NAND | 4-in NAND |
|---|---|---|---|---|
| **No Heat Extraction** | 2631K | 1711K | 1569K | 1367K |
| **Heat Extraction** | 384K | 374K | 368K | 364K |

As we can see from the results, due to the high density, long thermal paths, as well as surface scattering and confinement effects which reduce the thermal conductivity of thin nanowires, S3DC circuits can reach very high temperature when no fabric-level heat management component is applied.

With HEJ (one for each gate) and HDPP placed in the circuits, hot spot temperature reduces by up to 85%. Although we pessimistically assumed that all gate input / output wires provide no heat dissipation, critical temperature is still 384K, which is below the threshold temperature for modern microprocessors [14], and proves the intrinsic heat management fabric components to be effective.

*B. Performance, Power and Density Evaluation*

We have established an evaluation methodology based on extensive bottom-up simulations as shown in Figure VI for evaluating the performance, power and density of S3DC circuits. At the beginning, core fabric components including SB-ILC, transistors, and Ohmic contacts have been validated with 3D Sentaurus TCAD tools, which simulate both the process and device physics with nanoscale effects taken into account. These resulting device characteristics were analyzed by DataFit [15] with regression analysis and polynomial fits to acquire the mathematical expressions, based on which we built the behavioral HSPICE models. We also implemented benchmarks in detailed 3D physical layout, based on which the transistor-level HSPICE netlists are built and RC extractions are done following the Predictive Technology Model for interconnections [16] and using actual dimensions and materials. Then by using the previously built device models, we completed a physical level HSPICE simulation for circuit functionality validation. With this methodology, we ensure solid results accounting for manufacturing parameters, material considerations, circuit topology and wiring parasitics.

TCAD simulations of V-GAA Junctionless transistors show on-currents of 17 μA for n-type and 16 μA for p-type transistors, and a ~1e+5 on-off ratio; simulation results of SB-ILC have proved that it provides good Ohmic contacts between different doping regions of a nanowire.

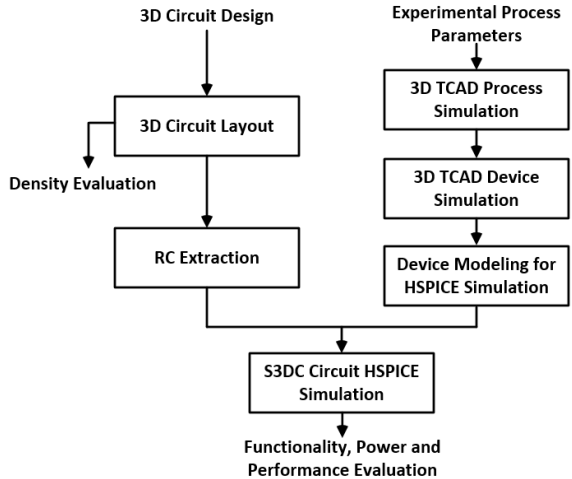

Figure VI. S3DC Evaluation Methodology

We have benchmarked 3D 4-bit and 16-bit array-based multipliers as well as a 3D 4-bit Wire Streaming Processor (WISP-4) [6] which are as yet manually designed (ongoing effort in our group is CAD-tooling development for S3DC) after which their functionality is validated. All these metrics were compared with 16-nm CMOS implementations with FinFETs, which are generated by a semi-custom design flow using CAD tools for synthesis, placement & routing and RC extraction.

TABLE II. BENCHMARK RESULTS

| | Metrics | CMOS | S3DC |
|---|---|---|---|
| **4-bit Multi** | Throughput (ops./sec.) | 4.97E+9 | 4.55E+9 (0.915x) |
| | Power (μW) | 172 | 33.9 (0.19x) |
| | Performance / Watt (ops./J) | 2.89E+13 | 1.34E+14 (4.64x) |
| | Density (mm$^{-2}$) | 2E+4 | 6.58E+5 (32.9x) |
| **16-bit Multi** | Throughput (ops./sec.) | 4.48E+9 | 4.57E+9 (1.02x) |
| | Power (μW) | 1.26E+4 | 1.29E+3 (0.102x) |
| | Performance / Watt (ops./J) | 3.56E+11 | 3.55E+12 (9.97x) |
| | Density (mm$^{-2}$) | 6.94E+2 | 2.75E+4 (39.6x) |
| **WISP-4 CPU** | Throughput (ops./sec.) | 4.31E+9 | 4.55E+9 (1.06x) |
| | Power (μW) | 886 | 186 (0.21x) |
| | Performance / Watt (ops./J) | 4.86+12 | 2.45E+13 (5.04x) |
| | Density (mm$^{-2}$) | 3.46E+3 | 9.43E+4 (27.3x) |

From the benchmark results we can see significant improvement in density (30-40X for circuits studied). S3DC also improves power efficiency (up to 10X performance per watt) while performance remains comparable. We expect large IC designs to fare even better due to S3DC's connectivity benefits, requiring much lesser number of repeaters due to a higher ratio of short interconnects vs medium and global wires.

Finally, as shown in Table II, density results are substantially better than what has been shown to be possible with state-of-the-art monolithic 3D approaches that were limited to benefits within 2X of CMOS in ideal case.

## IV. S3DC MANUFACTURING PATHWAY AND EXPERIMENTAL VALIDATION

In this section, the S3DC manufacturing pathway is introduced, and the manufacturing feasibility of S3DC fabric is discussed including highlighting related experimental demonstrations.

*A. S3DC Manufacturing Pathway*

The S3DC manufacturing pathway is shown with an example flow of building an S3DC Vertical Gate-All-Around (V-GAA) Junctionless transistor in Figure VII. As we can see, it is based on multi-layer material insertion to functionalize a uniform nanowire template. In S3DC one processes an IC as a single wafer in contrast to the parallel / monolithic 3D integration, which manufactures circuits in a layer-by-layer manner. Furthermore, S3DC fabric does not involve a selective doping process during nanowire template functionalizing; doping is necessary for fabricating each IC layer in monolithic 3D, which may harm the bottom layer circuits due to the high temperature dopant activation process. The S3DC manufacturing pathway allows the stacking of multiple components, such as transistors, contacts and metal routing structures within one doping layer of nanowires as we can see from previously shown circuit layouts. It also shifts the lithography precision requirement to material deposition, which is known to be controllable more precisely (and thus even could alleviate the lithography-imperfection-induced variations).

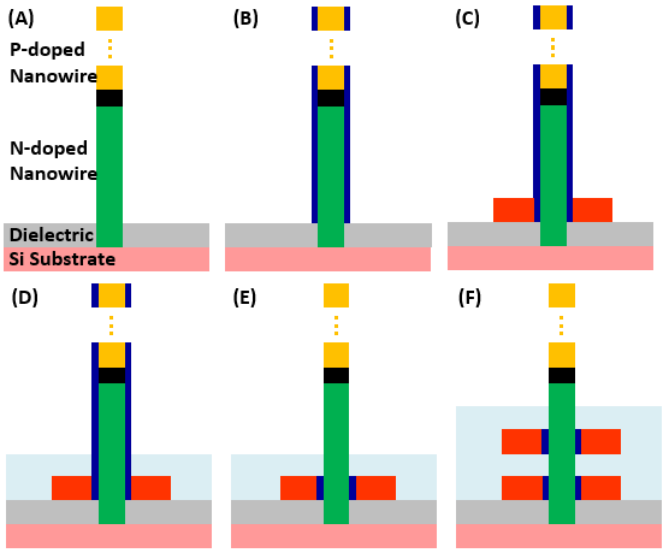

Figure VII. S3DC transistor fabrication: A). starting nanowire; the heavily-n-type-doped region for building n-type transistors; B). HfO$_2$ ALD for the gate dielectric formation; C). selective material deposition (TiN in this case) for gate electrode formation; D). insulator deposition and planarization; E). isotropic HfO$_2$ etching; F). more transistors sequentially stacked on one active layer.

*B. Experimental Demonstrations*

S3DC IC manufacturing generally includes two types of process: the uniform vertical nanowire template formation and multi-level selective material deposition. Therefore a validation of these two major steps is helpful to demonstrate the manufacturability of S3DC technology.

In order to form the template, firstly, one wafer containing several layers with p and n doping profiles is achieved by bonding individual p and n silicon wafers. Then vertical nanowires are achieved in the top-down manner by applying high aspect ratio anisotropic silicon etching to the prepared layered wafer. Every step during this template formation process has been demonstrated: wafer bonding technology has been widely used in current monolithic 3D integration and widely demonstrated [3]; vertical nanowire patterning can be achieved through processes such as Bosch Process [17], Inductively Coupled Plasma etching (~50:1 aspect ratio, 5nm dimension shown) [18], etc., and has been experimentally demonstrated in our group as shown in Figure VIII(A).

Following the nanowire patterning, multi-level selective material deposition functionalizes the template. Similarly with the deposition techniques in CMOS process, selective material deposition in S3DC manufacturing involves steps including lithography, planarization, deposition, lift-off, etc. Among these steps, planarization in S3DC is more challenging since the conventional Chemical Mechanical Polishing (CMP) process could cause structural damage to the vertical nanowires. Consequently, an alternative technique with etch-back on self-planarization material is used in S3DC. This technique planarizes the photoresist surface by coating thick self-planarizing resist (SU-8) layer to completely cover the nanowires and then etching the photoresist layer back to the desired thickness. This approach has been experimentally demonstrated in our group. All the other steps of material deposition can be done similarly to conventional CMOS manufacturing. Relying on the new planarization technique, precisely-controlled selective material depositions (various kinds of metal and oxide) in the S3DC nanowire template can be achieved and are shown in Figure VIII(B).

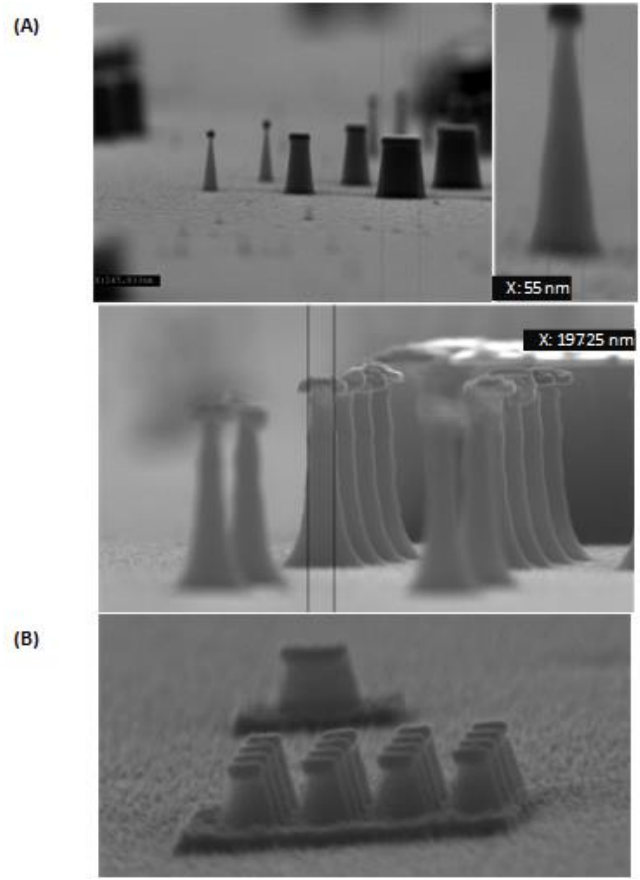

Figure VIII [19]. Cleanroom validations for S3DC manufacturability: A). vertical nanowire template demonstration: nanowires with different widths from 26nm-200nm (top figures) and with mostly uniform 197nm width and 1100nm height (bottom figures); B). metal-silicon contact as a demonstration of selective anisotropic metal deposition.

## V. COMPARISON BETWEEN S3DC AND OTHER 3D DIRECTIONS

As is shown in the fabric component section, the S3DC routing structures allow high flexibility in all three dimensions and thus feature unique connectivity approach. Wires in S3DC can "sneak" through the gaps between transistors and other functional elements along the nanowire grids, instead of being in the metal layers above the active layers. Figure IX shows the side-view inter-cell routing schematics of a random circuit implemented in different 3D technologies, which gives us an intuition on how the S3DC routing structures help to maintain good routability without using too many dedicated metal routing layers or seeing severe congestions, despite the tiny footprints of S3DC gates.

Other than the 3D interconnection structures, since S3DC allows stacking multiple transistors in one active layer, each transistor is directly accessible to adjacent components along all three axes, which also helps to shrink total wire length. The denser circuits decrease the inter-cell routing overhead as well. All these benefits have been summarized in Table III.

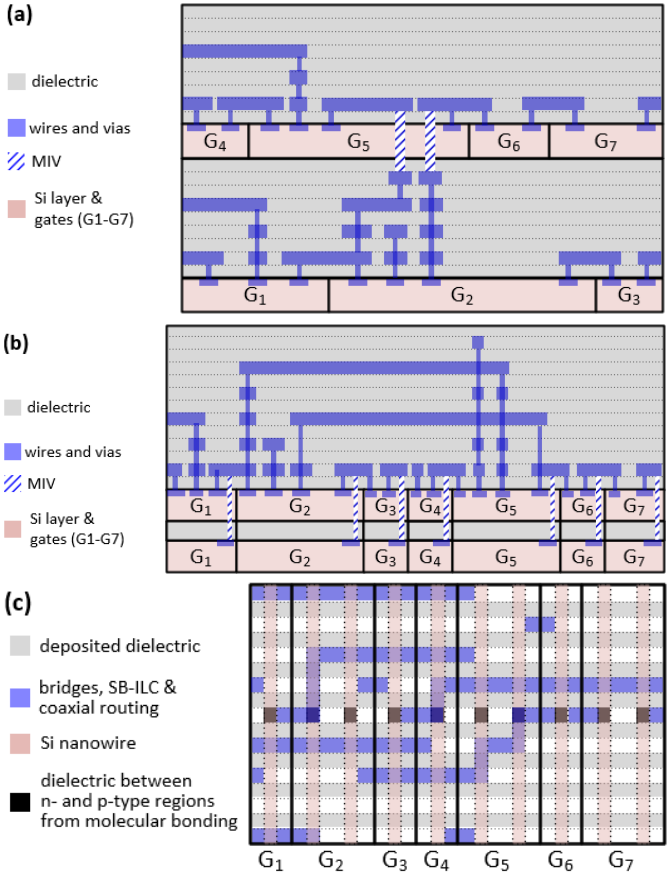

Figure IX. Inter-cell routing schematic of a random circuit in different fine-grained 3D IC technologies: A). gate-level monolithic 3D IC only adds limited inter-cell connectivity benefits through MIVs, and tends to use more metal layers since both top and bottom tiers need to be routed; B). transistor-level monolithic 3D IC improves intra-cell connectivity through MIVs while it follows most inter-cell routing conventions; pin and routing congestions are likely to happen due to smaller cell footprints; C). S3DC's flexible 3D routing allows most wiring done within active layers without severe congestions.

TABLE III. PARALLEL, MONOLITHIC 3D, VS. S3DC COMPARISON

|  | Parallel 3D | Monolithic 3D | True 3D w/ S3DC |
|---|---|---|---|
| Connectivity | Added connectivity from TSVs | Added connectivity from MIVs | Full 3D connectivity (vertical nanowire, Coaxial Routing and Bridges within one layer, SB-ILC between layers) |
| Granularity | Coarse-grained (limited by TSV alignment [1]) | Fine-grained (Cell- or transistor- level [5], layer-by-layer) | Truly fine-grained (transistor stacking within one active layer) |
| Process | Separate process for each layer | Layer -layer process, each layer doping | Processed as a single wafer |
| Potential | Up to ~2x (typically 40%) in density [20] assuming 2 stacked layers | Up to ~2x (typically 30-50%) in density [20] assuming 2 stacked layers | Up to 40X for initial circuits assuming two stacked gates Potential of 100X in density for very large designs. |

## VI. CONCLUSION

This article proposes the S3DC IC fabric, a vertically-composed fine-grained 3D IC technology alternative to the current 3D IC solutions. S3DC solves the 3D integration challenges with a new nanoscale fabric paradigm including a manufacturing pathway based on uniform 3D nanowire templates, novel 3D interconnection structures, and fabric-level heat management structures. The core new concepts have been validated with both detailed simulation and experiments. The yielded circuit-level benefits are found to be very significant vs. scaled CMOS and show great advantage vs. other types of 3D integration. Ongoing efforts include further S3DC experimental demonstrations of not only the key process steps and devices but also simple circuits as well as CAD tooling support.